# Improved Bounds and Schemes for the Declustering Problem[*]

Benjamin Doerr[†]    Nils Hebbinghaus[‡]    Sören Werth[§]

**Abstract**

The declustering problem is to allocate given data on parallel working storage devices in such a manner that typical requests find their data evenly distributed on the devices. Using deep results from discrepancy theory, we improve previous work of several authors concerning range queries to higher-dimensional data. We give a declustering scheme with an additive error of $O_d(\log^{d-1} M)$ independent of the data size, where $d$ is the dimension, $M$ the number of storage devices and $d-1$ does not exceed the smallest prime power in the canonical decomposition of $M$ into prime powers. In particular, our schemes work for arbitrary $M$ in dimensions two and three. For general $d$, they work for all $M \geq d-1$ that are powers of two. Concerning lower bounds, we show that a recent proof of a $\Omega_d(\log^{\frac{d-1}{2}} M)$ bound contains an error. We close the gap in the proof and thus establish the bound.

---

[*]supported by the DFG-Graduiertenkolleg 357 "Effiziente Algorithmen und Mehrskalenmethoden".

[†]Max–Planck–Institut für Informatik, Saarbrücken, Germany.
[‡]Max–Planck–Institut für Informatik, Saarbrücken, Germany.
[§]Institut für Informatik und Praktische Mathematik, Christian-Albrechts-Universität zu Kiel, Germany

# 1 Introduction

The last decade saw dramatic improvements in computer processing speeds and storage capacities. Nowadays, the bottleneck in data-intensive applications is the time needed to retrieve typically large amounts of data from external storage devices. One idea to overcome this obstacle is to distribute the data on disks of multi-disk systems so that it can be retrieved in parallel. Hopefully, this declustering reduces the retrieval time by a factor equal to the number of disks. The data allocation is determined by so-called declustering schemes. The schemes should allocate the data in such a manner that typical requests find their data evenly distributed on the disks.

We consider the problem of declustering uniform multi-dimensional data that is arranged in a multi-dimensional grid. There are many data-intensive applications that deal with this kind of data, especially multi-dimensional databases [CMA$^+$97, GM93, JRR99]. A range query $Q$ requests the data blocks that are associated with a hyper-rectangular subspace of the grid. Since we will not deal with syntactic issues of queries, we may identify a query with the set of requested block. In consequence, $|Q|$ denotes the number of requested blocks.

The response time of a query $Q$ is (proportional to) the maximum number of blocks of $Q$ that are assigned to the same disk (hence we assume identical disks). For an ideal declustering scheme for a system with $M$ disks, this would be $|Q|/M$ for all queries $Q$. As we will see, this aim cannot be achieved. The quality of a declustering scheme is measured by the worst case (over all queries $Q$) additive deviation of the response time from the ideal value $|Q|/M$.

The declustering problem for range queries is an intensively studied problem and a number of schemes [CBS03, PAGAA98, AP00, DS82, FB93] have been developed in the last twenty years. It was an important turning point when discrepancy theory was connected to declustering.

Before the use of discrepancy theory, no provable performance bounds were known for arbitrary dimension $d$. Such bounds existed only for a few rather restricted declustering schemes in two dimensions: For the scheme proposed in [CBS03], a proof for the average performance is given if the number $M$ of disks is a Fibonacci number. For the construction of the scheme in [AP00], $M$ has to be a power of 2.

A breakthrough was marked by noting that the declustering problem is a discrepancy problem. For the case $d = 2$, Sinha, Bhatia and Chen [SBC03] as well as Anstee, Demetrovics, Katona and Sali [ADKS00] developed declustering schemes for all $M$ and proved their asymptotically optimal behavior. The schemes of Sinha et al. [SBC03] are based on two dimensional low discrepancy point sets. They also give generalizations to arbitrary dimension $d$, but without bounds on the error.

Both papers show a lower bound of $\Omega(\log M)$ for the additive error of any declustering scheme in dimension two. The result of Anstee et al. [ADKS00] applies to latin square type colorings only, but their proof can easily be extended to the general case as well. Sinha et al. [SBC03] also claim a bound of $\Omega_d(\log^{\frac{d-1}{2}} M)$ for arbitrary dimension $d$, but their proof contains an error, that is critical for $d \geq 3$ (cf. Section 3).

The first non-trivial upper bounds for declustering schemes in arbitrary dimension were proposed by Chen and Cheng [CC02], who present two schemes for the $d$–dimensional declustering problem. The first one has an additive error of $O_d(\log^{d-1} M)$, but works only if $M = p^k$ for some $k \in \mathbb{N}$ and $p$ is a prime such that $d \leq p$. The second one works for arbitrary $M$, but the error increases with the size of the data. (Note that all other bounds stated in this paper are independent of the data size.)

**Our Results:** We work both on upper and lower bounds. For the upper bound, we present an improved scheme that yields an additive error of $O_d(\log^{d-1} M)$ for all values of $M$ (independent of the data size) and all $d$ such that $d \leq q_1 + 1$, where $q_1$ is the smallest factor in the canonical decomposition of $M$ into prime powers. This compares to the current best declustering scheme (with worst case additive error independent of the data size) due to Chen and Cheng [CC02] as follows. Its worst case additive error is of the same order of magnitude as ours, but has stronger restrictions on $M$. It works only if $M = p^k$ is a power of a prime and if $d \leq p$. Note that our scheme in the case $M = p^k$ requires only $d \leq p^k + 1$. Thus, in particular, for $M$ being a power of two our scheme can be used in every dimension $d \leq M + 1$, whereas the scheme in [CC02] only works for dimension $d = 2$. This and the fact that our scheme can be used for all $M$ in dimension 2 and 3, is useful from the viewpoint of application. After preparation of this paper and its conference version [DHW04], the journal version [CC04] of Chen and Cheng [CC02] was published. There, the quite strict limitations of [CC02] could be relaxed to the ones obtained in this paper.

We also show that the latin hypercube construction used by Chen and Cheng [CC02, CC04] and in our work is much better than proven there. Where they show that the final scheme has an error of at most $2^d$ times the one of the latin hypercube coloring, we show that both errors are the same.

For the lower bound, we present the first correct proof of the $\Omega_d(\log^{\frac{d-1}{2}} M)$ bound for dimension $d \geq 3$ and the $\Omega(\log M)$ bound for dimension $d = 2$. This is particularly interesting with regard to a recent result of Chedid [Che04]. There a declustering scheme is presented that works for $2^{2t}$ ($t \in \mathbb{N}$) disks in dimension $d = 2$. It is claimed that it has an additive error of at most 3.

## 2 Discrepancy Theory

In this section, we sketch the connection between the declustering problem and discrepancy theory.

### 2.1 Combinatorial Discrepancy

Recall that the declustering problem is to assign data blocks from a multi-dimensional grid to $M$ storage devices (disks) in a balanced manner. The aim is that range queries use all storage devices in a similar amount. More precisely, our grid is $V = [n_1] \times \cdots \times [n_d]$ for some positive integers $n_1, \ldots, n_d$.[1] A query $Q$ requests the data assigned to a *rectangle* (or *box*) $[x_1..y_1] \times \cdots \times [x_d..y_d]$ for some integers $1 \leq x_i \leq y_i \leq n_i$. We identify a query with the set of blocks it requests, i.e., $Q = [x_1..y_1] \times \cdots \times [x_d..y_d]$.

---

[1] We use the notations $[n] := \{1, 2, \ldots, n\}$ and $[n..m] := \{k \in \mathbb{N} \mid n \leq k \leq m\}$ for $n, m \in \mathbb{N}$, $n \leq m$.

We assume that the time to process a query is proportional to the maximum number of requested data blocks that are stored on a single device. We represent the assignment of data blocks to devices through a mapping $\chi : V \to [M]$. The processing time of the query $Q$ then is $\max_{i \in [M]} |\chi^{-1}(i) \cap Q|$, where, as usual, $\chi^{-1}(i) = \{v \in V : \chi(v) = i\}$. Clearly, no declustering scheme can do better than $|Q|/M$. Hence a natural performance measure is the additive deviation from this lower bound. We are interested in the worst-case behavior. Thus we are looking for declustering schemes such that $\max_Q \max_{i \in [M]} |\chi^{-1}(i) \cap Q|$ is small.

This makes the problem a combinatorial discrepancy problem in $M$ colors. Denote by $\mathcal{E}$ the set of all rectangles in $V$. Then $\mathcal{H} = (V, \mathcal{E})$ is a hypergraph. For a coloring $\chi : V \to [M]$, the discrepancy of a hyperedge $E \in \mathcal{E}$ with respect to $\chi$ is

$$\mathrm{disc}(E, \chi) := \max_{i \in [M]} \left| |\chi^{-1}(i) \cap E| - \tfrac{1}{M}|E| \right|,$$

the discrepancy of $\mathcal{H}$ with respect to $\chi$ is

$$\mathrm{disc}(\mathcal{H}, \chi) := \max_{i \in [M], E \in \mathcal{E}} \left| |\chi^{-1}(i) \cap E| - \tfrac{1}{M}|E| \right|,$$

and the discrepancy of $\mathcal{H}$ in $M$ colors is

$$\mathrm{disc}(\mathcal{H}, M) := \min_{\chi : V \to [M]} \mathrm{disc}(\mathcal{H}, \chi).$$

These definitions were introduced by Srivastav and the first author in [DS99, DS03] extending the well-known notion of combinatorial discrepancy to any number of colors. Similar notions were used by Biedl et al. [BČC$^+$02] and Babai, Hayes and Kimmel [BHK01]. For our purposes, only positive deviations have to be regarded ("too many blocks on one disk"). We adapt the multi-color discrepancy notion in the obvious way and define the positive discrepancy by

$$\mathrm{disc}^+(\mathcal{H}, \chi) := \max_{i \in [M], E \in \mathcal{E}} \left( |\chi^{-1}(i) \cap E| - \tfrac{1}{M}|E| \right),$$
$$\mathrm{disc}^+(\mathcal{H}, M) := \min_{\chi : V \to [M]} \mathrm{disc}^+(\mathcal{H}, \chi).$$

Clearly, we have $\tfrac{1}{M-1} \mathrm{disc}(\mathcal{H}) \le \mathrm{disc}^+(\mathcal{H}) \le \mathrm{disc}(\mathcal{H})$ for all hypergraphs $\mathcal{H}$. The first inequality follows from the fact that for every $E \in \mathcal{E}$ and every coloring $\chi : V \to [M]$ we have $\sum_{j \in [M]}(|\chi^{-1}(j) \cap E| - \frac{1}{M}|E|) = |E| - |E| = 0$. Summarizing the discussion above, we have the following.

**Theorem 1.** *The additive error of an optimal declustering scheme for range queries is* $\text{disc}^+(\mathcal{H}, M)$.

Since a central result of this paper are discrepancy bounds independent of the size of the grid, we usually work with the hypergraph $\mathcal{H}_N^d = ([N]^d, \mathcal{E}_N^d)$, $\mathcal{E}_N^d = \{\prod_{i=1}^d [x_i..y_i] \,|\, 1 \leq x_i \leq y_i \leq N\}$ for some sufficiently large integer $N$. Furthermore, we regard only the case that $M \geq 3$. For $M = 2$, a checkerboard coloring yields a declustering scheme with an additive error of $1/2$. We prove the following result.

**Theorem 2.** *Let $M \geq 3$ and $d \geq 2$ be integers and $q_1$ the smallest prime power in the canonical factorization of $M$ into prime powers. Then*

(i) $\text{disc}^+(\mathcal{H}_N^d, M) = O_d(\log^{d-1} M)$ *for* $d \leq q_1 + 1$, *independent of* $N \in \mathbb{N}$,

(ii) $\text{disc}^+(\mathcal{H}_N^d, M) = \Omega_d(\log^{\frac{d-1}{2}} M)$ *for* $N \geq M$,

(iii) $\text{disc}^+(\mathcal{H}_N^d, M) = \Theta(\log M)$ *for* $d = 2$.

## 2.2 Geometric Discrepancy

As mentioned before, the use of geometric discrepancies in the analysis of declustering problems in [SBC03, ADKS00] was a major breakthrough in this area. We refer to the recent book of Matoušek [Mat99] for both a great introduction and a thorough treatment of geometric discrepancies.

The geometric discrepancy problem is to distribute $n$ points evenly in a geometric setting. For our purposes, we regard discrepancies of point sets in $[0, 1]^d$ with respect to axis-parallel boxes. Such a box $R$ is the product $R = \prod_{i=1}^d [x_i, y_i)$ with $0 \leq x_i \leq y_i \leq 1$ for all $i \in [d]$. Our aim is that each box $R$ shall contain approximately $n \, \text{vol}(R)$ points, where $\text{vol}(R) = \prod_{i=1}^d (y_i - x_i)$ denotes the volume of $R$. Again, discrepancy quantifies the distance to a perfect distribution. The discrepancy of an $n$–point set $\mathcal{P}$ with respect to a box $R$ is defined by

$$D(\mathcal{P}, R) = \big||\mathcal{P} \cap R| - n \, \text{vol}(R)\big|,$$

the discrepancy of $\mathcal{P}$ with respect to the set $\mathcal{R}_d$ of all axis-parallel boxes is

$$D(\mathcal{P}, \mathcal{R}_d) = \sup_{R \in \mathcal{R}_d} |D(\mathcal{P}, R)|,$$

and the discrepancy of $\mathcal{R}_d$ for $n$-point sets is

$$D(n, \mathcal{R}_d) = \inf_{\substack{\mathcal{P} \subset [0,1)^d \\ |\mathcal{P}|=n}} D(\mathcal{P}, \mathcal{R}_d).$$

## 3  The Lower Bound

To prove our lower bounds, we use classical lower bounds for geometric discrepancies. Roth's [Rot54] famous lower bound for the $L_2$ discrepancy of the axis-parallel boxes immediately implies the following.

**Theorem 3 (Roth's lower bound).** *Let $d \geq 2$. There exists a constant $k > 0$ (depending on $d$) such that for any $n$–point set $\mathcal{P}$ in the unit cube $[0, 1)^d$, there is an axis-parallel box $R$ in $[0, 1)^d$ with*

$$D(\mathcal{P}, R) \geq k \log^{\frac{d-1}{2}} n.$$

It was Schmidt [Sch72] who came up with the sharp lower bound in two dimensions.

**Theorem 4 (Schmidt's lower bound).** *There is a constant $k > 0$ such that for any $n$–point set $\mathcal{P}$ in the unit square $[0, 1)^2$, there is an axis-parallel rectangle $R$ in $[0, 1)^2$ with*

$$D(\mathcal{P}, R) \geq k \log n.$$

The general idea in the proofs of the lower bound for declustering schemes in Sinha et al. [SBC03] and Anstee et al. [ADKS00] (for $d = 2$ only) is the following.

Any low-discrepancy $M$–coloring of $[M]^d$ has color classes of approximately $M^{d-1}$ vertices. By scaling, such a color class yields an $M^{d-1}$–point set $\mathcal{P}$ in $[0, 1)^d$. The lower bounds above give a box $R$ with polylogarithmic discrepancy. Round $R$ to a box $\bar{R}$ with corners in $\{0, \frac{1}{M}, \cdots, \frac{M-1}{M}, 1\}^d$ in such a way that $R \cap \mathcal{P} = \bar{R} \cap \mathcal{P}$. Then $R$ and $\bar{R}$ have similar volume and hence similar discrepancy. Rescaling $\bar{R}$ yields a hyperedge $\hat{R}$ with combinatorial discrepancy equal to the geometric one of $\bar{R}$.

The small, but crucial mistake in the proof of Sinha et al. [SBC03] is hidden in the transfer from the geometric discrepancy setting back to the combinatorial one. Unlike in dimension $d = 2$, rounding $R$ to $\bar{R}$ does not yield a constant change in the discrepancy in higher dimensions. The volume difference $|\text{vol}(R) - \text{vol}(\bar{R})|$ is still $O_d(\frac{1}{M})$. However, since the number of points is $M^{d-1}$, the change in the discrepancy can be of order $\Theta_d(M^{d-2})$. This is way too large for $d > 2$.

For this reason, a straight generalization of the proof of Anstee et al. [ADKS00] of the lower bound in two dimensions (as attempted in [SBC03]) is not possible. We solve this problem in the following way. Instead of looking at the whole $[M]^d$–grid, we focus on a *small* subgrid. This reduces the number of points, and hence the change in the discrepancy.

Here is an outline of the proof: Starting with an $M$–coloring of the $[M]^d$–grid we have to show the existence of a box with positive discrepancy of order $\Omega_d(\log^{\frac{d-1}{2}} M)$. We restrict the search to a small subgrid $[sM]^d$ (with $s$ a multiple of $\frac{1}{M}$) to avoid the above mentioned problems in the rounding process. The left part of Figure 1 depicts such an $[sM]^d$–subgrid. The crosses represent one color class. We choose this color class in such a way that it contains at least the average number of $s^d M^{d-1}$ vertices of $[sM]^d$. From this color class we get a set of points in the $[0, 1]^d$–cube by scaling. This can be seen in the middle part of Figure 1. Using the Theorem of Schmidt respectively Roth, we find a box $R$ (in the middle section of Figure 1) with large geometric discrepancy. We round this box to a box $\bar{R}$ containing the same points as $R$ but fitting to the grid lines stemming from the $[sM]^d$–grid. For the corresponding box $\hat{R}$ (the box with the continuous lines in the right section in Figure 1) in the $[sM]^d$–grid we estimate the discrepancy using the geometric discrepancy of the box $R$ and the relatively small change in the discrepancy caused by the volume difference between the boxes $R$ and $\bar{R}$. Should this large discrepancy be caused by a lack of vertices in one color, we get a lower bound for the positive discrepancy through the following observation. Either $\hat{R}$ or its complement in $[sM]^d$ has a positive discrepancy of the wanted order. Although the complement is not a box, it is the union of at most $2d$ boxes. Thus, at least one of these boxes has a positive discrepancy of order $\Omega_d(\log^{\frac{d-1}{2}} M)$.

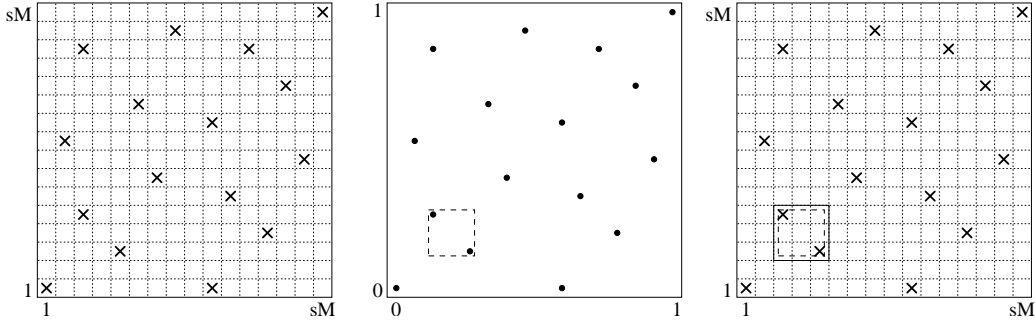

Figure 1: Construction of a box with large discrepancy.

*Proof of Theorem 2 (ii).* Clearly $\mathrm{disc}^+(\mathcal{H}_N^d, M) \geq \mathrm{disc}^+(\mathcal{H}_M^d, M)$ for all $N \geq M$. Hence, we can assume $N = M$. The proof is organized as follows. We first show a lower bound for the $M$–color discrepancy of $\mathcal{H}_M^d$. From this, we derive a lower bound for the positive $M$–color discrepancy of $\mathcal{H}_M^d$.

Let $\chi : [M]^d \to [M]$ be an $M$–coloring of $\mathcal{H}_M^d$. Choose an $s \in [M^{-\frac{d-2}{d-1}}, 2M^{-\frac{d-2}{d-1}}) \cap [0,1]$ such that $s$ is a multiple of $\frac{1}{M}$. Such an $s$ exists since $M^{-\frac{d-2}{d-1}} > \frac{1}{M}$. Without loss of generality, we may assume that $n := |\chi^{-1}(1) \cap [sM]^d| \geq s^d M^{d-1}$.

**Claim.** *There is a box $\hat{R} \subseteq [sM]^d$ such that*

$$\left| |\chi^{-1}(1) \cap \hat{R}| - \tfrac{1}{M}|\hat{R}| \right| = \Omega_d\left(\log^{\frac{d-1}{2}} M\right).$$

If $n \geq s^d M^{d-1} + \frac{k}{2}\left(\frac{1}{d-1}\right)^{\frac{d-1}{2}} \log^{\frac{d-1}{2}} M$, we clearly have $|n - s^d M^{d-1}| \geq \frac{k}{2}\left(\frac{1}{d-1}\right)^{\frac{d-1}{2}} \log^{\frac{d-1}{2}} M$. Therefore, we may assume

$$s^d M^{d-1} \leq n < s^d M^{d-1} + \tfrac{k}{2}\left(\tfrac{1}{d-1}\right)^{\frac{d-1}{2}} \log^{\frac{d-1}{2}} M. \qquad (1)$$

For every vertex $z = (z_1, z_2, \ldots, z_d) \in \chi^{-1}(1) \cap [sM]^d$ we define $x_z := \left(\frac{2z_1-1}{2sM}, \frac{2z_2-1}{2sM}, \ldots, \frac{2z_d-1}{2sM}\right)$. Let $\mathcal{P} := \{x_z \mid z \in \chi^{-1}(1) \cap [sM]^d\}$. Then $\mathcal{P}$ is an $n$-point set in the unit cube $[0,1)^d$. Estimating the cardinality of $\mathcal{P}$, we get $n \geq s^d M^{d-1} \geq M^{-\frac{(d-2)d}{d-1}} M^{d-1} = M^{\frac{-(d^2-2d)+(d-1)^2}{d-1}} = M^{\frac{1}{d-1}}$. By Theorem 3, there exists a box $R = \prod_{i=1}^d [x_i, y_i)$ in $[0,1)^d$ with

$$||R \cap \mathcal{P}| - n \operatorname{vol}(R)| \geq k \log^{\frac{d-1}{2}} n \geq k \left(\frac{1}{d-1}\right)^{\frac{d-1}{2}} \log^{\frac{d-1}{2}} M. \qquad (2)$$

Now we construct a box $\bar{R} = \prod_{i=1}^d [\bar{x}_i, \bar{y}_i)$ by rounding the $x_i$ and $y_i$ to the nearest multiple of $\frac{1}{sM}$. In case of ties, we round down. This ensures $\mathcal{P} \cap \bar{R} = \mathcal{P} \cap R$ as the following argument shows. Let $\left(\frac{2z_1-1}{2sM}, \frac{2z_2-1}{2sM}, \ldots, \frac{2z_d-1}{2sM}\right) \in \mathcal{P} \cap R$. This is equivalent to $x_i \leq \frac{2z_i-1}{2sM} < y_i$ for all $i \in [d]$. But this holds if and only if we have $\bar{x}_i \leq \frac{z_i-1}{sM}$ and $\bar{y}_i \geq \frac{z_i}{sM}$ for all $i \in [d]$, which is equivalent to $\left(\frac{2z_1-1}{2sM}, \frac{2z_2-1}{2sM}, \ldots, \frac{2z_d-1}{2sM}\right) \in \mathcal{P} \cap \bar{R}$.

We now quantify the effect of this rounding. The symmetric difference of $R$ and $\bar{R}$ is the union of $2d$ boxes such that all their side lengths are at most $1$ and one side length of each box is bounded from above by $\frac{1}{2sM}$ (this is due to the rounding process). Hence $|\operatorname{vol}(R) - \operatorname{vol}(\bar{R})| \leq 2d \frac{1}{2sM} = \frac{d}{sM}$. Using $s \leq 2M^{-\frac{d-2}{d-1}}$, we get

$$s^d M^{d-1} |\operatorname{vol}(R) - \operatorname{vol}(\bar{R})| \leq \frac{d}{sM} s^d M^{d-1} = d s^{d-1} M^{d-2} \leq d 2^{d-1}, \qquad (3)$$

an estimation needed below. Note that the choice of $s$ being small ensures that the effect of rounding is independent of $M$.

The combinatorial counterpart of $\bar{R}$ is the box

$$\hat{R} := \left\{ x \in [M]^d \,\Big|\, \left(\frac{2x_1-1}{2sM}, \ldots, \frac{2x_d-1}{2sM}\right) \in \bar{R} \right\}.$$

Hence,

$$|\chi^{-1}(1) \cap \hat{R}| = |\mathcal{P} \cap \bar{R}| = |\mathcal{P} \cap R|.$$

One also easily verifies that $|\hat{R}| = s^d M^d \operatorname{vol}(\bar{R})$. By construction,

$$\begin{aligned}
\left||\chi^{-1}(1) \cap \hat{R}| - \tfrac{1}{M}|\hat{R}|\right| &= \left||\mathcal{P} \cap \bar{R}| - s^d M^{d-1} \operatorname{vol}(\bar{R})\right| \\
&= \big||\mathcal{P} \cap R| - n \operatorname{vol}(R) + (n - s^d M^{d-1}) \operatorname{vol}(R) \\
&\quad + s^d M^{d-1} (\operatorname{vol}(R) - \operatorname{vol}(\bar{R})) \big| \\
&\geq \big||\mathcal{P} \cap R| - n \operatorname{vol}(R)\big| - |n - s^d M^{d-1}| \\
&\quad - s^d M^{d-1} |\operatorname{vol}(R) - \operatorname{vol}(\bar{R})|.
\end{aligned}$$

Observe that we bound the $M$–color discrepancy of $\mathcal{H}_M^d$ from below by the geometric discrepancy of the box $R$ minus two terms, comprising the fact that $n$ is not exactly $s^d M^{d-1}$ and the effect of rounding $R$ to $\hat{R}$. Hence by (1), (2) and (3),

$$\left||\chi^{-1}(1) \cap \hat{R}| - \tfrac{1}{M}|\hat{R}|\right|$$
$$\geq k\left(\tfrac{1}{d-1}\right)^{\frac{d-1}{2}} \log^{\frac{d-1}{2}} M - \tfrac{k}{2}\left(\tfrac{1}{d-1}\right)^{\frac{d-1}{2}} \log^{\frac{d-1}{2}} M - d2^{d-1}$$
$$= \Omega_d\left(\log^{\frac{d-1}{2}} M\right).$$

Thus, we have shown the claimed existence of a box $\hat{R} \subseteq [sM]^d$ with $\left||\chi^{-1}(1) \cap \hat{R}| - \tfrac{1}{M}|\hat{R}|\right| = \Omega_d(\log^{\frac{d-1}{2}} M)$. It remains to prove that this bound also holds for the positive discrepancy. To this end, let us assume that the discrepancy of the box $\hat{R}$ in color 1 is caused by a lack of vertices in color 1. Since $|\chi^{-1}(1) \cap [sM]^d| \geq s^d M^{d-1}$, the complement of $\hat{R}$ in $[sM]^d$ has at least the same discrepancy as $\hat{R}$, but caused by an excess of vertices in color 1. Though this complement is not a box, it is the union of at most $2d$ boxes. Therefore, one of these boxes has a positive discrepancy that is at least $\tfrac{1}{2d}$ times the discrepancy of $\hat{R}$ in color 1. $\square$

This last argument increases the implicit constant of the lower bound by a factor of $\tfrac{3^d}{2d}$ compared to the approach of Sinha et al. [SBC03].

We briefly show how to use the above to prove the $\Omega(\log M)$ bound for dimension $d = 2$. For this bound, two not completely satisfying bounds exist. Anstee et al. [ADKS00] only treated latin square type colorings of $[M]^2$ and posed it an open problem to extend their result to arbitrary colorings. The proof in [SBC03] does not have this restriction, but is not very precise, which in particular helped to hide the error for $d > 2$.

As a simple and clean proof we therefore propose the following: Use the same reasoning as in the case of arbitrary dimension $d \geq 2$, but apply Schmidt's lower bound instead of Roth's. The parameter $s$ can be choosen as 1. In dimension $d = 2$ we do not need small boxes, because the roundoff error has an effect on the discrepancy which is of order $O(1)$.

# 4 The Upper Bound

In this section, we present a declustering scheme showing our upper bound. As in previous work, we use low discrepancy point sets to construct the declustering scheme. In the following we use the notation of Niederreiter [Nie87]. For an integer $b \geq 2$, an *elementary interval* in base $b$ is an interval of the form $E = \prod_{i=1}^{d} \left[ a_i b^{-d_i}, (a_i + 1) b^{-d_i} \right)$, with integers $d_i \geq 0$ and $0 \leq a_i < b^{d_i}$ for $1 \leq i \leq d$. For integers $t, m$ such that $0 \leq t \leq m$, a $(t, m, d)$–*net* in base $b$ is a point set of $b^m$ points in $[0,1)^d$ such that all elementary intervals with volume $b^{t-m}$ contain exactly $b^t$ points.

Note that any elementary interval with volume $b^{t-m}$ has discrepancy zero in a $(t, m, d)$–net. Since any subset of an elementary interval of volume $b^{t-m}$ has discrepancy at most $b^t$ and any box can be packed with elementary intervals in a way that the uncovered part can be covered by $O_d(\log^{d-1} b^m)$ elementary intervals of volume $b^{t-m}$, the following is immediate:

**Theorem 5.** *A $(t, m, d)$–net $\mathcal{P}_{net}$ in base $b$ with $n = b^m$ points has discrepancy*
$$D(\mathcal{P}_{net}, \mathcal{R}_d) = O_d(\log^{d-1} n).$$

The central argument in our proof of the upper bound is the following result of Niederreiter [Nie87] on the existence of $(0, m, d)$–nets. From the view-point of application it is important that his proof is constructive. Admittedly, this construction is highly involved. We refer to the book of Niederreiter [Nie87] for the details.

**Theorem 6.** *Let $b \geq 2$ be an arbitrary base and $b = q_1 q_2 \ldots q_u$ be the canonical factorization of $b$ into prime powers such that $q_1 < \cdots < q_u$. Then for any $m \geq 0$ and $d \leq q_1 + 1$ there exists a $(0, m, d)$–net in base $b$.*

We use $(0, m, d)$–nets to construct an $M$–coloring of $\mathcal{H}_M^d$ in Lemma 7. For the definition of these colorings, we need the following special elements of $\mathcal{E}_M^d$: A set $\prod_{j=1}^{d} I_j \in \mathcal{E}_M^d$ is called a *row* of $[M]^d$ if there is an $i \in [d]$ with $I_i = [1..M]$ and $|I_j| = 1$ for all $j \neq i$. In Lemma 8 we use the $M$–coloring of $\mathcal{H}_M^d$ to construct an $M$–coloring of $\mathcal{H}_N^d$ with same discrepancy.

**Lemma 7.** Let $\mathcal{P}_{net}$ be a $(0, d-1, d)$–net in base $M$ in $[0,1)^d$. Then there is an $M$–coloring $\chi_M$ of $\mathcal{H}_M^d = ([M]^d, \mathcal{E}_M^d)$ such that all rows of $[M]^d$ contain every color exactly once[2] and

$$\operatorname{disc}(\mathcal{H}_M^d, \chi_M) \leq D(\mathcal{P}_{net}, \mathcal{R}_d).$$

*Proof.* The net $\mathcal{P}_{net}$ consists of $M^{d-1}$ points and all elementary intervals with volume $M^{-d+1}$ contain exactly one point. In particular, all subsets $\prod_{j=1}^d I_j$ of $[0,1)^d$ such that there is an $i \in [d]$ with $I_i = [0,1)$ and for all $j \neq i$ there exist $a_j \in [0..M-1]$ with $I_j = [\frac{a_j}{M}, \frac{a_j+1}{M})$, contain exactly one point.

We construct a coloring $\chi_M$ of $\mathcal{H}_M^d = ([M]^d, \mathcal{E}_M^d)$ corresponding to the set $\mathcal{P}_{net}$. Let $\hat{\mathcal{P}} := \left\{ x \in [M]^d \,\middle|\, \mathcal{P}_{net} \cap \prod_{i=1}^d [\frac{x_i-1}{M}, \frac{x_i}{M}) \neq \emptyset \right\}$. Then each row of $[M]^d$ contains exactly one point of $\hat{\mathcal{P}}$. We define the coloring $\chi_M : [M]^d \to [M]$ by $\chi_M(y, x_2, \ldots, x_d) = i$ for all $x = (x_1, x_2, \ldots, x_d) \in \hat{\mathcal{P}}$, $i, y \in [M]$ such that $y \equiv x_1 + (i-1) \mod M$. Hence $\hat{\mathcal{P}}$ receives color 1, color class 2 is obtained from shifting $\hat{\mathcal{P}}$ along the first coordinate and so on. This defines an $M$–coloring $\chi_M$ of $\mathcal{H}_M^d = ([M]^d, \mathcal{E}_M^d)$. Since each color class is constructed by shifting the first color class, each row of $\mathcal{H}_M^d$ contains every color exactly once. Thus, each whole row of $\mathcal{H}_M^d$ has discrepancy zero.

For this coloring it is sufficient to calculate $\max_{\hat{R} \in \mathcal{E}_M^d} \left| |\chi_M^{-1}(1) \cap \hat{R}| - \frac{1}{M}|\hat{R}| \right|$, because for each color $i \in [M]$ and each box $\hat{R} \in \mathcal{E}_M^d$ we get the same discrepancy for the box $\hat{R}'$, which is a copy of $\hat{R}$ shifted along the first dimension by $i-1$ and wrapped around perhaps, with respect to the color 1. If $\hat{R}'$ is wrapped around, it is the union of two boxes. Since whole rows have discrepancy zero, the discrepancy of those boxes is the same as the discrepancy of the box between them, and we have

$$\operatorname{disc}(\mathcal{H}_M^d, \chi_M) = \max_{\hat{R} \in \mathcal{E}_M^d} \left| |\hat{\mathcal{P}} \cap \hat{R}| - \frac{1}{M}|\hat{R}| \right|.$$

Let $\hat{R} = \prod_{i=1}^d [x_i..y_i]$ an arbitrary hyperedge of $\mathcal{H}_M^d$. The associated box in $[0,1)^d$ is $R = \prod_{i=1}^d \left[ \frac{x_i-1}{M}, \frac{y_i}{M} \right)$. Then $|\hat{\mathcal{P}} \cap \hat{R}| = |\mathcal{P}_{net} \cap R|$ and $|\hat{R}| = M^d \operatorname{vol}(R)$.

---
[2]Some authors call this a permutation scheme for $[M]^d$

Thus the combinatorial discrepancy of $\hat{R}$ equals the geometric one of $R$. We have

$$\left||\chi_M^{-1}(1) \cap \hat{R}| - \tfrac{1}{M}|\hat{R}|\right| = \left||\mathcal{P}_{net} \cap R| - M^{d-1}\operatorname{vol}(R)\right| \leq D(\mathcal{P}_{net}, \mathcal{R}_d).$$

Hence we get $\operatorname{disc}(\mathcal{H}_M^d, \chi_M) \leq D(\mathcal{P}_{net}, \mathcal{R}_d)$. $\square$

In the previous lemma we constructed an $M$–coloring for the $[M]^d$–grid with low discrepancy. We now extend this coloring to $[N]^d$–grids for arbitrary $N \in \mathbb{N}$. We do this by plastering the $[N]^d$–grid with copies of the $[M]^d$–grid coloring.

**Lemma 8.** *Let $\chi_M$ be an $M$–coloring of $\mathcal{H}_M^d$ such that all rows of $[M]^d$ contain every color exactly once and $\chi$ a coloring of $\mathcal{H}_N^d$ defined by $\chi(x_1, \ldots, x_d) = \chi_M(y_1, \ldots, y_d)$ with $x_i \equiv y_i \mod M$ for $i \in [d]$, $x_i \in [N]$, $y_i \in [M]$. Then*
$$\operatorname{disc}(\mathcal{H}_N^d, \chi) = \operatorname{disc}(\mathcal{H}_M^d, \chi_M).$$

*Proof.* The proof is organized in the following way. Pick an arbitrary box in the $[N]^d$–grid. Using the fact that whole rows in the $[M]^d$–grid coloring $\chi_M$ have discrepancy zero, we can ignore all of the box except its corners. By construction, these corners can all be found in one common $[M]^d$–subgrid. Since whole rows (in the $[M]^d$–grid coloring $\chi_M$) have discrepancy zero, taking complements in each dimension does not alter the discrepancy. We thus obtain a box in the $[M]^d$–grid that has the same discrepancy as the original box.

Let $\hat{R} = \prod_{i=1}^d [x_i..y_i]$ be an arbitrary hyperedge of $\mathcal{H}_N^d$. For all $i \in [d]$ there exist unique $\widetilde{x}_i, \widetilde{y}_i \in [M]$ with $x_i \equiv \widetilde{x}_i \pmod{M}$ respectively $y_i \equiv \widetilde{y}_i \pmod{M}$. If $\widetilde{x}_i \leq \widetilde{y}_i$, we set $\bar{x}_i := \widetilde{x}_i$ and $\bar{y}_i := \widetilde{y}_i$. Otherwise we set $\bar{x}_i := \widetilde{y}_i + 1$ and $\bar{y}_i := \widetilde{x}_i - 1$. We define the rectangles

$$\begin{aligned}
\hat{R}_l &:= [\widetilde{x}_1..M] \times [x_2..y_2] \times \ldots \times [x_d..y_d], \\
\hat{R}_r &:= [1..\widetilde{y}_1] \times [x_2..y_2] \times \ldots \times [x_d..y_d], \\
\hat{R}_0 &:= [\bar{x}_1..\bar{y}_1] \times [x_2..y_2] \times \ldots \times [x_d..y_d].
\end{aligned}$$

Using the fact that whole rows have discrepancy zero and the fact, that the coloring $\chi$ is invariant under shifts with multiples of $M$ in any dimension, we

get for all $i \in [M]$

$$\left||\hat{R} \cap \chi^{-1}(i)| - \tfrac{1}{M}|\hat{R}|\right| = \left||\hat{R}_l \cap \chi^{-1}(i)| - \tfrac{1}{M}|\hat{R}_l| + |\hat{R}_r \cap \chi^{-1}(i)| - \tfrac{1}{M}|\hat{R}_r|\right|$$
$$= \left||\hat{R}_0 \cap \chi^{-1}(i)| - \tfrac{1}{M}|\hat{R}_0|\right|.$$

Thus, $\text{disc}(\hat{R}, \chi) = \text{disc}(\hat{R}_0, \chi)$. Applying this successively in every coordinate, we get

$$\text{disc}(\hat{R}, \chi) = \text{disc}(\prod_{i=1}^d [\bar{x}_i..\bar{y}_i], \chi) = \text{disc}(\prod_{i=1}^d [\bar{x}_i..\bar{y}_i], \chi_M).$$

This completes the proof. $\square$

Lemma 8 is a remarkable improvement of Theorem 4.2 in [CC02], where $\text{disc}(\mathcal{H}_N^d, \chi) \leq 2^d \text{disc}(\mathcal{H}_M^d, \chi_M)$ is shown. Note that this reduces the implicit constant in the upper bound by factor of $2^d$.

It remains to show that the upper bound in Theorem 2 follows from Lemma 7 and Lemma 8.

*Proof of Theorem 2(i).* Let $M \geq 3$ and $d \geq 2$ be positive integers and $d \leq q_1 + 1$, where $q_1$ is the smallest prime power in the canonical factorization of $M$ into prime powers. Theorem 6 provides a $(0, d-1, d)$–net $\mathcal{P}_{net}$ in base $M$ in $[0, 1)^d$. Using Lemma 7, we get an $M$–coloring $\chi_M$ of $\mathcal{H}_M^d$ such that all rows contain each color exactly once and $\text{disc}(\mathcal{H}_M^d, \chi_M) \leq D(\mathcal{P}_{net}, \mathcal{R}_d)$. With Lemma 8 and Theorem 5, we have $\text{disc}(\mathcal{H}_N^d, M) \leq D(\mathcal{P}_{net}, \mathcal{R}_d) = O_d(\log^{d-1} M)$. $\square$

## 5 Conclusion

We gave lower and upper bounds for the declustering problem of range queries to higher-dimensional grids. This paper contains the first complete proof of the lower bound $\Omega_d(\log^{\frac{d-1}{2}} M)$ for arbitrary values of $M$ and $d$.

We proposed a declustering scheme that has an additive error of $O_d(\log^{d-1} M)$ with the sole condition that $d \leq q_1 + 1$, where $q_1$ is the smallest prime power in the canonical factorization of $M$ into prime powers. This improves the former best declustering schemes of Chen and Cheng [CC02], where either bounds depend on the data size $N^d$ or $M = p^t$ and $p \geq d$ was required for some prime $p$ and $t \in \mathbb{N}$. Furthermore, Lemma 8 improves the analysis of Chen and Cheng [CC02, CC04] of the discrepancy of latin square colorings by a factor of $2^{-d}$.

The natural problem arising from this work is to close the gap between the lower and upper bound. However, this is probably a very hard one. The reason is that the corresponding problem for geometric discrepancies of boxes is extremely difficult. Closing the gap between the $\Omega_d(\log^{\frac{d-1}{2}} n)$ lower and the $O_d(\log^{d-1} n)$ upper bound for $D(n, \mathcal{R}_d)$ was baptized 'the great open problem' already in Beck and Chen [BC87]. Since then no further progress has been made for the general problem. Note that in the proof of a slight improvement due to Baker [Bak99] recently a serious error was found, so that the result was withdrawn by the author [reported by József Beck, Oberwolfach Seminar on Discrepancy Theory and Applications, March 2004].